\documentclass[11pt]{article}
\usepackage{amsfonts}
\usepackage{amssymb}
\def\bsh{\backslash}
\def\bdt{\dot \beta}
\def\adt{\dot \alpha}

\newfont{\bbbold}{msbm10 scaled \magstep1}

\def\bbM{\mbox{\bbbold M}}

\newfont{\goth}{eufm10 scaled \magstep1}

\def\gl{\mbox{\goth l}}

\def\gp{\mbox{\goth p}}

\def\gs{\mbox{\goth s}}

\def\gu{\mbox{\goth u}}

\def\a{\alpha}
\def\b{\beta}

\def\d{\delta}\def\D{\Delta}

\def\l{\lambda}
\def\m{\mu}

\def\p{\pi}

\def\S{\Sigma}
\def\t{\tau}

\def\be{\begin{equation}}\def\ee{\end{equation}}
\def\bea{\begin{eqnarray}}\def\eea{\end{eqnarray}}
\def\ba{\begin{array}}\def\ea{\end{array}}

\def\del{\partial}

\def\str{\rm str}
\def\xz{\times}

\def\del{\partial}

\let\la=\label

\def\bd{\begin{document}}
\def\ed{\end{document}}
\def\bea{\begin{eqnarray}}
\def\eea{\end{eqnarray}}
\def\ba{\begin{array}}
\def\ea{\end{array}}
\def\ft#1#2{{\textstyle{{\scriptstyle #1}\over {\scriptstyle #2}}}}
\def\fft#1#2{{#1 \over #2}}
\newcommand{\eq}[1]{(\ref{#1})}
\def\eqs#1#2{(\ref{#1}-\ref{#2})}
\def\det{{\rm det\,}}
\def\tr{{\rm tr}}
\newcommand{\ho}[1]{$\, ^{#1}$}
\newcommand{\hoch}[1]{$\, ^{#1}$}
\newcommand{\tamphys}{\it\small Center for Theoretical Physics,
Texas A\&M University, College Station, TX 77843, USA} 
\newcommand{\newton}{\it\small Isaac Newton Institute for Mathematical
Sciences, Cambridge, UK} 
\newcommand{\kings}{\it\small Department of Mathematics, King's College,
London, UK} 
\newcommand{\lapp}{\it\small LAPP, Annecy, France}

\newcommand{\auth}{\large P.S. Howe\hoch{a},
C. Schubert\hoch{b}, E. Sokatchev\hoch{b} and P.C. West\hoch{a}} 
\begin{document}\input feynman

\thispagestyle{empty}

\hfill{KCL-MTH-99-41} 

\hfill{LAPTH-746/99} 

\hfill{\today}

\vspace{20pt} 

\begin{center}
{\Large{\bf Explicit construction of nilpotent covariants in $N=4$ 
SYM}} \vspace{30pt} 

\auth

\vspace{15pt} 

\begin{itemize}
\item [$^a$] \kings
\item [$^b$] {\it\small Laboratoire d'Annecy-le-Vieux de Physique
Th{\'e}orique\footnote{UMR 5108 associ{\'e}e {\`a} 
l'Universit{\'e} de Savoie} LAPTH, Chemin de Bellevue, B.P. 110, 
F-74941 Annecy-le-Vieux, France} 
\end{itemize}

\vspace{60pt} 
{\bf Abstract} 

\end{center}

Some aspects of correlation functions in $N=4$ SYM are discussed.
Using $N=4$ harmonic superspace we study two and three-point
correlation functions which are of contact type and argue that
these contact terms will not affect the non-renormalisation
theorem for such correlators at non-coincident points. We then
present a perturbative calculation of a five-point function at two
loops in $N=2$ harmonic superspace and verify that it reproduces
the derivative of the previously found four-point function with 
respect to the coupling. The calculation of this four-point function
via the five-point function turns out to be significantly simpler
than the original direct calculation.
This  calculation also provides an 
explicit construction of an $N=2$  component of an $N=4$ 
five-point nilpotent covariant that violates $U(1)_Y$ symmetry. 

{\vfill\leftline{}\vfill

\pagebreak \setcounter{page}{1}
 \section{Introduction}

The superconformal Ward identities play a central r{\^o}le in the
study of superconformal field theories, and in particular, in
$N=4$ Yang-Mills theory. These Ward identities can be conveniently
expressed in harmonic superspaces - either $N=4$ on-shell harmonic
superspace \cite{hhh} which has the advantage that the
gauge-invariant operators in short multiplets are represented by
single-component analytic superfields, or $N=2$ off-shell harmonic
superspace \cite{hh} in which, although the $N=4$ Yang-Mills
multiplet decomposes into the $N=2$ Yang-Mills multiplet plus a
hypermultiplet, one can carry out perturbation theory calculations
whilst maintaining manifest $N=2$ supersymmetry.

 Using the harmonic superspace approach to $N=4$ SYM and motivated
by earlier works \cite {n} which indicated that the Ward identities for correlation functions of constrained superfield operators
in superconformal quantum field theories 
are stronger than one might naively expect
some interesting results were found. In particular
\cite{hw1,hw2}, the SYM field strength is a covariantly
analytic superfield $W$ carrying no indices from which one can build a
set of analytic gauge-invariant operators by tracing products of $W$.
The members of this set are in one-to-one correspondence with the
Kaulza-Klein multiplets of IIB supergravity on $AdS_5\times S^5$
\cite{af} and  the Ward identities for correlation functions of
operators of this type are easy to solve for in terms of prefactors
times functions of superconformal invariants, largely due to the fact
that the fields carry no indices. Since there are no three-point
superinvariants for the superspaces we are considering, it is possible
to obtain the functional form of three- (and two-) point functions
exactly \cite{hw1,hsw}, although the Ward identities do not determine
the dependence of the coefficients on the coupling. For four and more
points, the Ward identities, when  combined with analyticity, do
put constraints on the functions of superinvariants that can arise
\cite{eetall,toaapear}, although  these constraints do not
seem to be enough to determine completely the $N=2$ correlators that
contain four harmonic matter fields of charge two contrary to the 
conjecture made in \cite{hw1,hw2}.
However, it is not ruled out that this line of argument cannot be used
to show that other correlators can be found explicitly
\footnote{Indeed, in a
paper in preparation \cite{ehssw} we shall use harmonic superspace
arguments of the above type to show that extremal correlators of
analytic operators are free. This is 
in accordance with the recent results of
\cite{dhf} in AdS and also with the results 
obtained in \cite{bk} for
the one-loop and instanton contributions
to the corresponding correlators on the
field theory side.}.
\par
To
make further progress, therefore, it seems that in general some 
additional input
is required. A   field-theoretic trick one can use is to derive a
relation  between
$n$- and $(n+1)$-point functions by differentiating the  path integral
representation of the
$n$-point function with  respect to the coupling. In the present
context we shall refer to  the resulting equation as the reduction
formula; it was  first applied to $N=4$ SCFT in \cite{ken1}. An
important aspect of  this formula is that the $(n+1)$-point function
includes an  integration over the point of insertion of the
 Lagrangian.
\par
One application of the reduction formula is to use
  the known explicit form of all three- and
four-point  superconformal invariants to prove the
non-renormalisation  theorem for two- and three-point functions
\cite{ehw}. We  mentioned above that the Ward identities fix the form
of these functions  but not their dependence on the coupling although,
for the Green's  function of three supercurrents, there is an argument
\cite{hsw},  based on the absence of counterterms beyond one loop in
$N=4$  conformal supergravity \cite{hst} which implies that this
Green's  function depends trivially on the coupling (see also
\cite{dzf}).  The result of \cite{ehw} extends this
non-renormalisation theorem  to three-point functions of short
multiplet operators with  arbitrary charges. In a sense it is an
expression of the $U(1)_Y$  ``bonus'' symmetry first proposed  in
\cite {fera}, and advocated in \cite{ken1}.
 Although this  is  not
a true symmetry of interacting
$N=4$ Yang-Mills theory,  it nevertheless seems to be a symmetry of a
large class of  superinvariants and through them of
$n$-point functions of short  operators with $n\leq 4$ \cite{ken1,ehw}.

These results on two- and three-point functions are in accord with
the conjectured relation between $N=4$ SYM and IIB supergravity on
an $AdS_5\xz S^5$ background \cite{m}. However, it was emphasised
in \cite{skenderis} that contact terms can arise in the field
theory and that such terms can in principle have an effect in
the reduction formula because of the presence of an integrated
insertion. Contact terms have been observed in two-point functions
at two loops in the $N=2$ harmonic superspace formalism in
\cite{hsw} and studied in more detail in $N=1$ superfields in
\cite{stonybrook}. The authors of \cite{skenderis} were
particularly interested in the effect of such terms on anomalies,
but the contact terms found at two loops are actually consistent
with the Ward identities. Since the superconformal anomaly is
related to the divergences of $N=4$ conformal supergravity, and
since, as we remarked above, there are no such divergences beyond
one loop, it follows that the contact terms which could
potentially arise in the reduction formula should be consistent
with the superconformal Ward identities.

In this article we begin by discussing contact term solutions to
the $N=4$ superconformal Ward identities using $N=4$ superfields.
For two points we find that there are  covariant contact terms for
short multiplet operators with arbitrary charges. For three points
we then  find all possible  contact terms some of
which are nilpotent  and some which
are not. The latter are not so important because they cannot
contribute in the reduction formula. For the former we find
that there only exists a
 solution for the case of three supercurrents. A
consequence of this is that, if the formula is to be valid for
contact terms as well as at non-coincident points, it should be
the case that no contact terms should occur for two-point
functions with higher charges than the supercurrent. It
has to be admitted that this is not easy to verify directly in
 perturbation theory due to difficulties
that arise in defining the graphs for a small number of points.
However, one can certainly say that the only three-point contact
nilpotent covariant does not affect the two-point
non-renormalisation theorem; inserted in the reduction formula it
simply reproduces the two-point contact term for two
supercurrents. Although we have not investigated the situation at
four points in as much detail, we have identified the four-point
nilpotent contact covariant which gives rise to the three-point
one using the reduction formula, and we present an argument that suggests that
there  are no additional four-point contact covariants which could
interfere with the  proof of the three-point non-renormalisation
theorem. It should be borne in mind, however, that analyticity can be violated by harmonic delta-function terms and the effects of this are difficult to analyse in the $N=4$ formalism as it is on-shell. This caveat also applies to the $N=4$ version of the reduction formula (see below) which cannot be derived directly in contrast to the $N=2$ version. We are reassured by $N=2$ perturbative calculations that these difficulties should not affect our results.
\par
Another possible use of the reduction formula is to obtain
information about four-point correlators by first trying to guess
or compute the corresponding five-point ones. It should be
emphasised that, starting with five points, one can have nilpotent
superconformal covariants of the non-contact type. In the
reduction formula, after the integration over the insertion point,
they can become non-nilpotent. In order to reproduce the known
four-point correlators in this way, one has to assume the
existence of a five-point covariant violating $U(1)_Y$ invariance
\cite{ken1}. In \cite{ehw} it was shown that such covariants can
only be of the nilpotent type and their expression to lowest order in
fermions was given up to a multiplicative non-nilpotent function.
\par
In the second part of the paper we are able to 
construct explicitly an $N=2$ component of such an $N=4$ covariant by
calculating a certain five-point function   at two loops in
$N=2$ harmonic superspace.  This confirms the existence of the
nilpotent covariant and gives its form in detail. The five-point
correlator is related by the reduction formula to the correlator of four $N=2$ hypermultiplet bilinear (or charge two) composites. In
\cite{eetal} we  derived expressions for these correlators in terms of three functions of the spacetime variables
$A_1, A_2$  and $A_3$. The first two of these come out as functions of
the two  independent spacetime conformal cross-ratios. In fact, $A_1,
A_2$  are expressed in terms of the one-loop scalar box integral.
However, $A_3$ is given by a generic two-loop integral for which
conformal invariance is far from obvious. Subsequently, in
\cite{eetall}, we used the superconformal Ward identities combined
with harmonic analyticity to find a relation between $A_3$ and the
other two. This enabled us to show that $A_3$ is expressed in
terms of the same one-loop scalar box integral. Thus we could
verify the conformal invariance of the entire four-point
amplitude. This result was then confirmed by a numerical study of
the integral formula for $A_3$. The calculation we present here is
a two-loop computation at five points with four charge two
hypermultiplet composites and one Yang-Mills composite
$\mbox{Tr}\; W^2$. This fifth operator introduces a chiral point
which one is to integrate over in the Intriligator formula. As
expected, we reproduce the two-loop four-point function in a way
which makes the simplified form of $A_3$ immediately apparent.
Furthermore the calculation confirms the existence of a five-point
$N=4$ nilpotent superconformal invariant which is not invariant
under $U(1)_Y$. The calculation also gives direct support to the
assertion that contact terms do not make any significant
difference to the reduction formula, since none are required in
this case.


\section{Contact covariants}

We briefly recall the analytic superspace formalism. $N=4$
analytic superspace $\bbM$ has coordinates
\begin{equation}
X^{A A'}=\left( \ba{ll} x^{\a\adt} & \l^{\a a'} \\ \p^{a\adt} &
y^{a a'} \ea\right) \end{equation} where each lower case index can
take on 2 values. The even coordinates  $x$ and $y$ are
coordinates for complex spacetime and the internal space
$S(U(2)\xz U(2))\bsh SU(4)$ respectively. The odd coordinates $\l$
and $\p$ number 8 in all, half the number of odd coordinates of
$N=4$ super Minkowski space. An infinitesimal superconformal
transformation takes the form
\begin{equation}
\d X= V X=B + AX + XD + XCX \label{sct} \end{equation} where each
of the parameter matrices is a $(2|2)\xz (2|2)$ supermatrix and
where
\begin{equation}
\d g=\left( \ba{ll} -A & B \\ -C & D\ea\right) \in \gs\gl(4|4)\;.
\end{equation} One can show that the central elements in the superalgebra
$\gs\gl(4|4)$ do not  act on $\bbM$ so that one really has an
action of the superalgebra $\gp\gs\gl(4|4)$.

>From \eq{sct} one can read off the vector fields for each of the
parameters. They divide into translational ($B$), linear ($A,D$)
and quadratic ($C$) types. The translations are ordinary spacetime
translations, half of the $Q$-supersymmetry transformations and
translations in the internal $y$ space, which is locally the same
as spacetime. The corresponding vector fields are

\begin{equation}
V_{AA'}= {\del\over\del X^{AA'}} \label{trans}\;. \end{equation}

The linearly realised symmetries are Lorentz transformations
($SL(2)\xz SL(2)$ in complex spacetime) and dilations, a
corresponding set of internal transformations, the other half of
the $Q$-supersymmetries and half of the $S$-supersymmetries. The
$SL(2)$ transformations are handled in the usual way so that we do
not need to write them down. The vector fields generating
dilations ($D$) and internal dilations ($D'$) are

\begin{eqnarray} V(D) &=& x^{\a\adt}\del_{\a\adt} + {1\over2}(\l^{\a
a'}\del_{\a a'}+\p^{a\adt}\del_{a\adt})\;, \\ V(D') &=& y^{a
a'}\del_{a a'} + {1\over2}(\l^{\a a'}\del_{\a
a'}+\p^{a\adt}\del_{a\adt})\;. \label{dil} \end{eqnarray}

The vector fields generating linearly realised $Q$-supersymmetry
are

\begin{eqnarray} V(Q)^a_{\a} &=& \p^{a\adt}\del_{\a\adt} +y^{a a'}\del_{\a
a'}\;,\\ V(Q)^{a'}_{\adt} &=& \l^{\a a'}\del_{\a\adt}-y^{a
a'}\del_{a\adt}\;, \label{linq} \end{eqnarray}

while those generating linearly realised $S$-supersymmetry are

\begin{eqnarray} V(S)^{\a}_a &=& x^{\a\adt}\del_{a \adt} + \l^{\a a'}\del_{a
a'}\;,\\ \label{lins1} V(S)^{\adt}_{a'} &=& x^{\a\adt}\del_{\a
a'}-\p^{a\adt}\del_{a a'}\;. \label{lins2} \end{eqnarray}

The remaining supersymmetry transformations are the non-linearly
realised $S$-supersymmetries generated by

\begin{eqnarray} V(S)^{\adt a}&=& x^{\b\adt}\p^{a\bdt}\del_{\b\bdt} +
x^{\b\adt}y^{a b'}\del_{\b b'} -\p^{b\adt}\p^{a\bdt}\del_{b\bdt}
-\p^{b\adt} y^{a b'}\del_{b b'}\;,\\ V(S)^{a' \a}&=& -\l^{\b
a'}x^{\a\bdt}\del_{\b\bdt} -\l^{\b a'}\l^{\a b'}\del_{\b b'} +
y^{b a'}x^{\a\bdt}\del_{b\bdt} + y^{b a'}\l^{\a b'}\del_{b b'}\;.
\label{quads} \end{eqnarray}

Finally, we have conformal boosts ($K$) and internal conformal
boosts ($K'$) generated by

\begin{eqnarray} V(K)^{\a\adt} &=& x^{\b\adt} x^{\a\bdt}\del_{\b\bdt}+
x^{\b\adt} \l^{\a b'}\del_{\b
b'}+\p^{b\adt}x^{\a\bdt}\del_{b\bdt}+\p^{b\adt}\l^{\a b'}\del_{b
b'}\;,\\ V(K')^{a a'} &=& \l^{\b a'}\p^{a\bdt}\del_{\b\bdt} +
\l^{\b a'}y^{a b'}\del_{\b b'} + y^{b a'}\p^{a\bdt}\del_{b\bdt} +
y^{b a'} y^{a b'}\del_{b b'}\;. \label{boost} \end{eqnarray}

The gauge-invariant operators in short multiplets in $N=4$ SYM are
$A_q=\mbox{Tr}\; (W^q)$ where $W$ is  the $N=4$ SYM field strength
tensor which takes its values in the Lie algebra $\gs\gu(N_c)$ of
the gauge group. These operators transform as
\begin{equation}
\d A_q=V A_q + q\D A_q \end{equation} where $\D=\str (A+XC)$. A
correlation function of such operators
\begin{equation}
G(X_1,\ldots, X_n)=<A_{q_1}(X_1)\ldots A_{q_n}(X_n)>
\end{equation} should satisfy the Ward identity
\begin{equation}
\sum_{i=1}^n (V_i + q_i\D_i) G=0\;. \end{equation}

We define a contact covariant to be a Green's function which
satisfies the Ward identities and which is local in the sense that
it involves at least one spacetime delta-function. Translational
symmetries imply immediately that any $n$-point Green's function
depends only on $n-1$ coordinate differences, $X_{12},X_{23}\ldots
X_{(n-1)n}$ where $X_{12}=X_1-X_2$. Furthermore, assuming that
there are no delta-functions in the internal space (which would be
inconsistent with analyticity), linear $Q$-supersymmetry can be
used to eliminate a further set of odd coordinates. Thus, after
imposing translational and linear $Q$-supersymmetry we find that a
Green's function may be taken to depend on $n-2$ $\l$ and $\p$
coordinates of the form

\begin{eqnarray} \l_{123}&=&\l_{12} y_{12}^{-1} -\l_{23} y_{23}^{-1}\;,\\
\p_{123}&=&y_{12}^{-1}\p_{12} - y_{23}^{-1}\p_{23}\;.
\end{eqnarray}

as well as $n-1$ $y$ differences and $n-1$ modified $x$
differences, $\hat x_{12},\ldots$, with

\begin{equation}\label{hat4} \hat x_{12}=x_{12}-\l_{12}y_{12}^{-1}\p_{12}\;.
\end{equation}

In the preceding three equations we have suppressed the indices as
the quantities involved in each expression are arranged in such a
way that matrix multiplication is natural provided that the
indices on $y^{-1}$ are taken to be a pair of subscripts in the
order $a' a$.

From the foregoing it follows that any two-point Green's function,
whether contact or not, cannot depend explicitly on the odd
coordinates. For a contact two-point function spacetime dilations
fix the dependence on $\hat x_{12}$ to be of the form of powers of
the d'Alembertian acting on the delta-function. Internal dilations
then give the dependence on $y_{12}$ and so we arrive at the
candidate two-point functions

\begin{equation}
<A_q(1) A_q(2)>\sim (y_{12})^{2q} \square^{2q-4}\d(\hat x_{12})\;.
\end{equation}

It is now a straightforward exercise to check that the expression
on the right-hand side does indeed satisfy the remaining Ward
identities. We note that the case of $q=2$, i.e. the two-point
function of two supercurrents, is the example previously
encountered in perturbation theory \cite{hsw,stonybrook}, although
the complete expression for the entire multiplet has not been
derived before to our knowledge.

We now turn to three-point functions. We are primarily interested
in nilpotent three-point covariants as they can contribute to
the reduction formula. We remind the reader that this reads, in
$N=4$ superspace,

\begin{equation}
{\del\over \del\t}<A_{q_1}\ldots  A_{q_n}>={1\over \t_2}\int d\m
<TA_{q_1}\ldots  A_{q_n}>
\la{2pt}
\end{equation}

where the integral is over the point of the inserted supercurrent
$T(=A_2)$. The measure $d\m$ involves an integral over the
internal coset and a fermionic integral over $\l$, i.e. $d\m\sim
d^4x\,du\,d^4\l$. Using linear $S$-supersymmetry one can show that
a three-point nilpotent Green's function with non-coincident
points cannot depend on $\l_{123}$ or $\p_{123}$ and hence must
vanish completely. However, the same is not true in the presence
of delta-functions. Consider the following expression which
satisfies the translational, $Q$-supersymmetry and dilational Ward
identities for three supercurrents,

\begin{equation}
<T(1)T(2)T(3)>\sim (\l_{123})^4 (y_{12})^4 (y_{23})^4 \d(\hat
x_{12})\d(\hat x_{23})\;. \label{ttt} \end{equation}

The variation of $(\l_{123})^4$ under the first linear
$S$-supersymmetry \eq{lins1} is zero, but under the second linear
$S$ transformation the variation depends linearly on $\hat x_{12}$
and $\hat x_{23}$. Explicitly,

\begin{equation}
V(S)^{\adt}_{a'}(\l_{123})^4=(\hat
x_{12}^{\a\adt}(y_{12}^{-1})_{a'a}-\hat
x_{23}^{\a\adt}(y_{23}^{-1})_{a'a}) (\l_{123}^3)^a_{\a}\;.
\end{equation}

However, this variation vanishes for the entire right-hand side of
\eq{ttt} due to the presence of the delta-functions. To complete
the proof that the proposed three-point function satisfies all the
Ward identities, it is sufficient to check the quadratic
$S$-supersymmetry transformations \eq{trans} since the
supersymmetries generate the entire super Lie algebra.  After some
algebra one can verify straightforwardly that the function does
transform in the right way under this symmetry.

We remark that integrating the right-hand side of \eq{ttt} over
point 1, one recovers the functional form of the two-point contact
term given above with $q=2$. In other words, the two- and
three-point contact terms for supercurrent Green's functions are
related to each other by the reduction formula, and the contact
three-point function therefore has no effect on the two-point
function at non-coincident points. We note further that, although
the three-point function given above does not seem to be symmetric
under the interchange of any two points, in fact it is. This is
partly due to the presence of the delta-functions and partly
because the $\l^4$ term can be written in the form

\begin{equation}
(\l_{123})^4={(\l_{123})^4\over (y_{13})^4} (y_{13})^4\;.
\end{equation}

The $(y_{13})^4$ factor here combines with the other two
$y$-factors to give a symmetrical expression while
${(\l_{123})^4\over (y_{13})^4}$ is symmetrical by itself.

At first sight it might seem that the above Green's function could
be generalised to higher charges by the inclusion of appropriate
d'Alembertians. However, this is  not  the case.
Consider a general, nilpotent contact three-point function, $G$,
of the type that can contribute in the reduction formula. It will
have the form

\begin{equation}
G=(\l_{123})^4F(\hat x_{12},\hat x_{23},y_{12}, y_{23})\;.
\end{equation}

Under the second linear $S$-supersymmetry the fermionic factor
will contribute a term

\begin{equation}
V(S)^{\adt}_{a'} G\sim (\hat
x_{12}^{\a\adt}(y_{12}^{-1})_{a'a}-\hat
x_{23}^{\a\adt}(y_{23}^{-1})_{a'a}) (\l_{123}^3)^a_{\a} F\;.
\end{equation}

However, it is easily seen that this term cannot be cancelled by
terms arising from the variations of the $y$'s or the $\hat x$'s
because the fermion structures are different. Therefore the above
term must vanish identically. The unique permissible $\hat
x$-dependence of $F$ which ensures this is the product of two
delta-functions, $\d(\hat x_{12})\d(\hat x_{23})$.  However, this
$\hat x$ structure demands the $y$ structure appearing in \eq{ttt}
and so we conclude that this contact covariant is the only one of
its type.

One can also consider four-point contact terms that only have a
$\lambda^4$ multiplied by a non-nilpotent factor. By suitable
labelling we may choose this to be
$\lambda_{123}^4$. Using
S-supersymmetry and repeating the argument leading to \eq{tttt}
we find that the non-nilpotent factor must contain $\delta (\hat
x_{12})\delta (\hat x_{23}) $. Using the
reduction formula we see that such a term cannot lead to a three-point function that
contains no  delta functions. As a result we conclude that
contact terms  cannot invalidate the non-renormalisation theorem for
two- and three-point functions shown in
\cite {ehw}.

It is straightforward to construct a sequence of  nilpotent contact
covariants for an arbitrary number of points which are related by the
reduction formula. For four points the covariant is

\be
<T(1)T(2)T(3)T(4)>\sim (\l_{123})^4 (\l_{234})^4 (y_{12})^4 (y_{23})^4 (y_{34})^6 \d(\hat x_{12})\d(\hat x_{23})\d(\hat x_{34})
\la{tttt}
\ee

The proof that this satisfies the appropriate Ward identity  is
straightforward; one simply observes that the right-hand side is
almost a product of two three-point functions of the type of \eq{ttt};
in fact, morally it is the product of two such functions divided by a
non-nilpotent two-point contact function of the type given in \eq{2pt}
with $q=2$. Using this, one can show in a few lines that the
four-point Ward identity is indeed satisfied by \eq{tttt}.
\par
This construction can be extended to an arbitrary number of  points
straightforwardly. The contact covariant for $n$ $T$'s is simply:

\be
<T(1)\ldots T(n)>\sim \prod_{i=1}^{n-2} (\l_{(i(i+1)(i+2)})^4
\prod_{i=1}^{n-1} \big((y_{i(i+1)})^4 \d(\hat x_{i(i+1)})\big)
\la{nt}
\ee

This sequence of terms is clearly related to each other by the
reduction
formula and thus finally to the non-nilpotent two-point
function \eq{2pt} with $q=2$. One can also  show that all contact
terms which are not nilpotent and have no derivatives acting on the
delta functions can only have the form
\be
(y_{12}^2)^2\ldots (y_{n n-1}^2)^2\delta(\hat x_{12})\ldots
\delta (\hat x_{n n-1})
\ee
This can be expressed as a product of two-point functions and
so corresponds to  a disconnected Greens function. It should be
straightforward to extend this result to include derivatives on the
delta functions.
\par
We shall now argue that all contact terms which are not
disconnected   and do not have derivatives on delta functions  are of
the form of equation (29). Given any non-nilpotent contact term we
can integrate over  the variable associated with a given leg to
produce a contact term with one less external leg. Repeating this
process and assuming one does not get zero one will arrive at a
non-nilptotent contact term that must be of the form of the above
equation. If we further assume that Greens functions that begin as
connected do not become disconnected then we would conclude that the
only connected Greens contact Greens functions are those that
lead by repeated integration to the non-nilpotent two-point
function and so are as given in equation (29).
If we assume that this
result also holds for contact terms with derivatives on the delta functions
we can conclude that all the connected contact Green's functions are
fixed by a single coefficient. The correlation functions of the
supercurrent can be generated from an effective supergravity action
obtained by coupling
$N=4$ SYM to a background supergravity and  integrating over the
Yang-Mills fields.
The contact terms must then arise from only one term in this
effective action which is superconformally invariant  in four
dimensions.
The unique superconformal function of the supergravity fields is
the $N=4$ conformal supergravity action. Differentiating this with
respect to the fields of this multiplet and setting them equal to
their flat space values should then give the contact covariants
described above.


\section{Two-loop calculation}

In this section we shall carry out a perturbative $N=2$ 
calculation at two loops which explicitly demonstrates how the 
reduction formula works. Our main aim will be to reproduce the 
four-point correlator of hypermultiplet bilinears from refs. 
\cite{eetal,eetall} as the integral of a five-point one. The 
latter is obtained by inserting the $N=2$ SYM Lagrangian into the 
four-point correlator. It provides a more explict form of the 
nilpotent  five-point superconformal covariant which was 
consturucted to lowest order in \cite {ehw}. This term violates 
the $U(1)_Y$  symmetry of ref. \cite{ken1}. 

\subsection{$N=4$ SYM in terms of $N=2$ harmonic superfields}

The absence of an off-shell formulation of $N=4$ SYM theory does 
not allow one to do perturbation theory calculations in a 
manifestly $N=4$ covariant way. The best one can do is reformulate 
the theory in terms of off-shell $N=2$ harmonic superfields and 
then apply the existing Feynman graph technique for such 
superfields. 

The two $N=2$ ingredients of the $N=4$ SYM theory are the $N=2$ 
SYM multiplet and the $N=2$ matter (hyper)multiplet. Both of them 
can be described as superfields in the Grassmann (G-)analytic 
superspace \cite{hh} with coordinates 
$x^{\alpha\dot\alpha}_A,\theta^{+\alpha}, 
\bar\theta^{+\dot\alpha},u^\pm_i$. Here $u^\pm_i$ are the harmonic 
variables which form a matrix of $SU(2)$ and parametrise the 
sphere $S^2\sim SU(2)/U(1)$. A harmonic function $F^{(q)}(u^\pm)$ 
of $U(1)$ charge $q$ is a function of $u^\pm_i$ invariant under 
the action of the group $SU(2)$ (which rotates the index $i$ of 
$u^\pm_i$) and homogeneous of degree $q$ under the action of the 
group $U(1)$ (which rotates the index $\pm$ of $u^\pm_i$). Such 
functions have infinite harmonic expansions on $S^2$ whose 
coefficients are $SU(2)$ tensors (multispinors). The superspace is 
called G-analytic since it only involves half of the Grassmann 
variables, the $SU(2)$-covariant harmonic  projections 
$\theta^{+\alpha} = u^+_i\theta^{i\alpha},\ 
\bar\theta^{+\dot\alpha} = u^+_i\bar\theta^{i\dot\alpha}$. 

In this framework the hypermultiplet is described by a G-analytic 
superfield of charge $+1$, $q^+(x_A,\theta^+,\bar\theta^+,u)$ (and 
its conjugate $\tilde q^+(x_A,\theta^+,\bar\theta^+,u)$ where 
$\tilde{}$ is a special conjugation on $S^2$ preserving 
G-analyticity). Note that this $N=2$ supermultiplet cannot exist 
off shell with a finite set of auxiliary fields \cite{nogo}. This 
only becomes possible if an infinite number of auxiliary fields 
(coming from the harmonic expansion on $S^2$) are present. On 
shell these  auxiliary fields are eliminated by the harmonic 
(H-)analyticity condition (equation of motion) 
\begin{equation}\label{EMo}
  D^{++}q^+ = 0\;.
\end{equation}
Here $D^{++}$ is the harmonic derivative on $S^2$ (the raising 
operator of the group $SU(2)$ realised on the $U(1)$ charges, 
$D^{++}u^+=0,\; D^{++}u^-=u^+$). In the G-analytic superspace it 
becomes a supercovariant operator involving spacetime derivatives: 
\begin{equation}\label{D++}
  D^{++} = u^{+i}{\partial\over\partial u^{-i}}
-4i\theta^{+\alpha}\bar\theta^{+\dot\alpha} {\partial\over\partial 
x^{\alpha\dot\alpha}_A} \;. 
\end{equation}
The field equation (\ref{EMo}) can be derived from an action given 
by an integral over the G-analytic superspace: 
\begin{equation}\label{6.7.1}
S_{\mbox{\scriptsize HM}} = -\int 
dud^4x_Ad^2\theta^+d^2\bar\theta^+\; \tilde q^{+}D^{++}q^+\;. 
\end{equation}
This action is real (with respect to the $\tilde{}$ conjugation) 
which can be seen by integrating $D^{++}$ by parts. In this sense 
the action (\ref{6.7.1}) resembles the Dirac action for fermions, 
although the superfield $q^+$ is bosonic.

The SYM gauge potential is introduced by covariantising the action 
(\ref{6.7.1}) with respect to a Yang-Mills group with G-analytic 
parameters $\lambda(x_A,\theta^+,\bar\theta^+,u)$. To this end one 
replaces the harmonic derivative in (\ref{6.7.1}) by the following 
covariant one: 
\begin{equation}
D^{++}\rightarrow \nabla^{++}=D^{++} + 
igV^{++}(x_A,\theta^+,\bar\theta^+,u) 
\end{equation}
where $g$ is the gauge coupling constant. The gauge potential is 
described by a real ($\widetilde {V^{++}} =V^{++}$) G-analytic 
superfield of charge $+2$ (equal to the charge of $D^{++}$). The 
matter and gauge superfields are subject to the usual gauge 
transformations: 
\begin{equation}\label{6.7.5}
{q^+}' = e^{ig\lambda}q^+\,, \ \ \ {V^{++}}' = -{i\over 
g}e^{ig\lambda}D^{++}e^{-ig\lambda} + 
e^{ig\lambda}V^{++}e^{-ig\lambda}\;, 
\end{equation}
so that the covariantised action (\ref{6.7.1}) 
\begin{equation}\label{HMcov}
  S_{\mbox{\scriptsize HM/SYM}} = -\int
dud^4x_Ad^2\theta^+d^2\bar\theta^+\; \tilde q^{+}\nabla^{++}q^+ 
\end{equation}
is indeed gauge invariant. 

The gauge invariant action for $V^{++}$ can be written down in 
terms of the gauge field strength $W(x_L,\theta^{i\alpha})$. 
Unlike the G-analytic potential, $W$ is a (left-handed) chiral 
superfield which is harmonic-independent, $\D^{++}W=0$. It can be 
expressed as a power series in $V^{++}$ \cite{Zup}: 
\begin{equation}\label{WV}
  W = {i\over 4} u^+_iu^+_j \bar D^i_{\dot\alpha}\bar D^{j\dot\alpha}
\sum^\infty_{n=1} \int du_1\ldots du_n\; {(-ig)^{n} V^{++}(u_1) 
\ldots V^{++}(u_n) \over (u^+u^+_1)(u^+_1u^+_2) \ldots (u^+_nu^+)} 
\end{equation}
where $(u^+_mu^+_n)\equiv u^{+i}_mu^+_{ni}\;$. The SYM action is 
then given by the chiral superspace integral \footnote{In fact, 
there exists an alternative form given by the right-handed chiral 
integral $\int d^4x_Rd^4\bar\theta\; {\rm Tr}\;\bar W^2$. In a 
topologically trivial background the two forms are equivalent (up 
to a total derivative).} 
\begin{equation}\label{SYMact}
  S_{\mbox{\scriptsize N=2 SYM}} =
{1\over 4g^2}\int d^4x_Ld^4\theta\;  {\rm Tr}\;W^2\;. 
\end{equation}
The details of how to fix the gauge and introduce ghosts can be 
found in \cite{hsgr}. 

When the hypermultiplet matter is taken in the adjoint 
representation of the gauge group, the two actions (\ref{HMcov}) 
and (\ref{SYMact}) describe the $N=4$ SYM theory, 
\begin{equation}\label{N4sym}
  S_{\mbox{\scriptsize N=4 SYM}} =  S_{\mbox{\scriptsize N=2 SYM}} +
 S_{\mbox{\scriptsize HM/SYM}}\;.
\end{equation}
As mentioned earlier, the main advantage of the $N=2$ formulation 
is the possibility to quantise the theory in a straightforward way 
\cite{hsgr}.

\subsection{The reduction formula in $N=2$}

The main aim of our perturbative calculation is to explicitly show 
how the general formula used by Intriligator in ref. \cite{ken1} 
works. It relates the correlation function of $n$ composite 
operators to an $(n+1)$-point one where the extra point is 
obtained by inserting the $N=4$ SYM Lagrangian. Here we shall 
derive this formula in the context of the $N=2$ harmonic 
superspace formulation of the $N=4$ theory. 

Consider a set of $n$ composite gauge invariant operators ${\cal 
O}_a$, $a=1,\ldots,n$, each made out of $r_a$ hypermultiplets 
$\tilde q^+$ and $s_a$ hypermultiplets $q^+$, $$ {\cal O}_a 
=(\tilde q^+)^{r_a}(q^+)^{s_a}\ .$$ Their correlator is given by 
the functional integral 
\begin{equation}\label{corr}
G_n = {\rm Tr}\;\langle{\mathcal O}_1\ldots{\mathcal 
O}_n\rangle = {1\over Z}\int {\mathcal D}q{\mathcal D}V\; 
e^{iS_{N=4\; SYM}} \; {\mathcal O}_1\ldots{\mathcal O}_n \;. 
\end{equation}

Now we want to differentiate equation (\ref{corr}) with respect to 
the coupling constant $g$. By inspecting the two ingredients 
(\ref{HMcov}) and (\ref{SYMact}) of the action (\ref{N4sym}), one 
sees that, after the change of variables 
\begin{equation}\label{chvar}
  V^{++}  \ \rightarrow \ {1\over g}V^{++}
\end{equation} in the functional integral,
the only dependence on $g$ is given by the overall factor $g^{-2}$ 
in the $N=2$ SYM part (\ref{SYMact}) of the action. Thus, we find 
\begin{eqnarray}
 {\partial G_n\over\partial g} &=&
{1\over Z}\int {\mathcal D}q{\mathcal D}V\; e^{iS_{N=4\; SYM}} 
\;{\partial (iS_{\mbox{\scriptsize N=2 SYM}})\over\partial g} \; 
{\mathcal O}_1\ldots{\mathcal O}_n \nonumber\\ 
  &=& -{2i\over  g} \int_{n+1}\langle
{\mathcal O}_1\ldots{\mathcal O}_n{1\over 4g^2}{\rm 
Tr}\;W^2_{n+1}\rangle\label{ourI}\\ 
 &=& -{2i\over  g} \int_{n+1}\langle
{\mathcal O}_1\ldots{\mathcal O}_n{\mathcal L}_{\mbox{\scriptsize 
N=2 SYM}}(n+1)\rangle\;. \nonumber 
\end{eqnarray}
Note that throughout the derivation we have used the 
gauge-invariant SYM Lagrangian instead of the gauge-fixed one. 
This is possible since, on the one hand, the composite operators 
${\mathcal O}$ are gauge invariant and on the other, the 
difference between the two forms of the gauge action amounts to a 
gauge (or BRST) transformation. So, this formula relates the 
$n$-point correlator of composite hypermultiplet operators to the 
$(n+1)$-point one obtained by inserting the $N=2$ SYM Lagrangian 
(without the matter part).\footnote{In its original version 
\cite{ken1} the formula involves a complex coupling constant 
$\tau$. This corresponds to including the topological part of the 
SYM action with a separate parameter $\theta$. Here we only 
consider a background of trivial topology, so our $g$ is real.} 

Intriligator's proposal was to use the formula (\ref{ourI}) to try 
to learn something about the $n$-point function by first 
predicting (or computing) the $(n+1)$-point one. The first half of 
the present paper was devoted to the possibility of predicting 
such correlators based on their superconformal properties. Now we 
shall undertake a direct calculation of the right-hand side of eq. 
(\ref{ourI}). We will deal with the correlation functions of four 
(two) bilinear composite operators made out of hypermultiplets and 
a fifth (third) bilinear representing the insertion of the SYM 
Lagrangian into the four (two)-point correlator. After integrating 
over the insertion point, we will recover the known results for 
the four (two)-point correlators of hypermultiplet bilinears. The 
five-point correlator (before integration) is an example of a 
nilpotent superconformal invariant preserving harmonic 
analyticity, but violating the $U(1)_Y$ invariance of ref. 
\cite{ken1}. 

\subsection{Graphs and Feynman rules}

We will be interested in the five-point correlator 
\begin{equation}\label{5pc}
  \langle(\tilde q^+(1))^2(q^+(2))^2(\tilde
q^+(3))^2(q^+(4))^2{1\over 4g^2}(W(5))^2\rangle\;. 
\end{equation}
We want to perform the computation at the lowest non-trivial level 
of perturbation theory, i.e., at two loops. For this reason we do 
not need to consider non-Abelian vertices and can restrict 
ourselves to the minimal coupling SYM/HM from eq. (\ref{HMcov}). 
The non-trivial graph topologies relevant to the computation are 
shown in Figure 1:  \vskip 1in 
\begin{center}
\begin{picture}(42000,7000)(0,-4500)

\drawline\fermion[\E\REG](0,0)[10000] \global\advance\pmidx by 
-400 \global\Yone=-1500 \put(\pmidx,\Yone){a} 
\global\advance\pmidx by 400 
\drawline\gluon[\N\CENTRAL](\pmidx,\pmidy)[6] 
\put(\pmidx,\pmidy){\circle*{1000}} \global\Xone=\gluonlengthy 
\drawline\fermion[\W\REG](\gluonbackx,\gluonbacky)[5000] 
\drawline\fermion[\E\REG](\gluonbackx,\gluonbacky)[5000] 
\drawline\fermion[\S\REG](\pbackx,\pbacky)[\Xone] 
\drawline\fermion[\N\REG](0,0)[\Xone] 

\drawline\fermion[\E\REG](13000,0)[10000] \global\advance\pmidx by 
-400 \put(\pmidx,\Yone){b} 
\drawline\fermion[\N\REG](\pbackx,\pbacky)[\Xone] 
\drawline\fermion[\W\REG](\pbackx,\pbacky)[10000] 
\drawline\fermion[\S\REG](\pbackx,\pbacky)[\Xone] 
\global\Xtwo=\pmidx \global\Ytwo=\pfronty \startphantom 
\drawline\gluon[\NE\FLIPPED](\pmidx,\pmidy)[3] \stopphantom 
\global\Ythree=\gluonlengthy \global\negate\Ythree 
\global\advance\Ytwo by \Ythree 
\drawline\gluon[\NE\FLIPPED](\Xtwo,\Ytwo)[3] 
\put(\pmidx,\pmidy){\circle*{1000}} 

\startphantom \drawloop\gluon[\N 5](26000,0) \stopphantom 
\global\Xfive=\loopfrontx \global\negate\Xfive 
\global\advance\Xfive by \loopbackx \global\advance\Xfive by 
-10000 \global\divide\Xfive by 2 
\drawline\fermion[\E\REG](26000,0)[10000] \global\advance\pmidx by 
-400 \put(\pmidx,\Yone){c} 
\drawline\fermion[\N\REG](\pbackx,\pbacky)[\Xone] 
\drawline\fermion[\W\REG](\pbackx,\pbacky)[10000] 
\global\advance\pmidy by -2650 \put(\pmidx,\pmidy){\circle*{1000}} 
\global\advance\pfrontx by \Xfive 
\drawloop\gluon[\S5](\pfrontx,\pfronty) 
\drawline\fermion[\N\REG](26000,0)[\Xone]

\global\advance\Yone by -1500 \put(15800,\Yone){Figure 1} 
\end{picture}
\end{center}

They have been obtained from the corresponding four-point graphs 
(see \cite{eetal} for details about the four-point calculation) by 
inserting the $N=2$ SYM linearised Lagrangian $W^2$ into each of 
the gluon lines. This amounts to replacing the gluon propagator 
\begin{center}
  \begin{picture}(0,3000)
  \drawline\gluon[\E\CENTRAL](-12000,0)[5]
  \global\advance\pfrontx by -800
  \put(\pfrontx,0)1
  \global\advance\pbackx by 500
  \put(\pbackx,0)2
  \global\advance\pbackx by 5000
  \put(\pbackx,0){$\langle V^{++}(1)V^{++}(2)\rangle$}
  \end{picture}
\end{center}
\vspace{5mm} by the product of modified propagators 
\begin{center}
  \begin{picture}(0,3000)
  \drawline\gluon[\E\CENTRAL](-15000,0)[7]
  \put(\pmidx,200){\circle*{1000}}
  \global\advance\pfrontx by -800
  \put(\pfrontx,0)1
  \global\advance\pbackx by 500
  \put(\pbackx,0)2
  \put(\pmidx,1200)3
  \global\advance\pbackx by 5000
  \put(\pbackx,0){$\langle V^{++}(1)W(3)\rangle
  {1\over 4g^2}\langle W(3)V^{++}(2)\rangle\ .$}
  \end{picture}
\end{center}
 \vspace{5mm}
The modified SYM propagator $\langle W(1)V^{++}(2)\rangle$ has one 
chiral end (the field strength $W(x_{1_L},\theta_1)$) and one 
G-analytic end (the SYM potential 
$V^{++}(x_{2_A},\theta^+_{2},\bar\theta^+_{2},u_2)$). One way to 
construct it is to take the standard SYM propagator $\langle 
V^{++}(1)V^{++}(2)\rangle$ in the Feynman gauge and convert the 
G-analytic end 1 to a chiral one using the linearised version of 
the expression (\ref{WV}) of the field strength $W$ in terms of 
the potential $V^{++}$. However, it is easier to guess the form of 
this mixed chiral-G-analytic object based only on its dimension 
and supersymmetry properties. We recall that the natural $N=2$ 
superspace for describing (left-handed) chiral objects is the 
chiral one with coordinates $x^{\alpha\dot\alpha}_L,\theta^{\alpha 
i} $, and that for G-analytic objects is the Grassmann-analytic 
harmonic superspace with coordinates 
$x^{\alpha\dot\alpha}_A,\theta^{+\alpha},\bar\theta^{+\dot\alpha},u^\pm_i$. 
\footnote{The subscripts of $x_L$ and $x_A$ refer to the 
appropriate bases in superspace where chirality or G-analyticity 
become manifest.} The transformation rules of these coordinates 
under $N=2$ supersymmetry are: 
\begin{equation}
  \begin{array}{lll}
   \mbox{Chiral superspace:}& &\mbox{G-analytic superspace:} \\
  & &\\
  \delta x^{\alpha\dot\alpha}_L = -4i\theta^{\alpha
i}\bar\epsilon^{\dot\alpha}_i & & \delta x^{\alpha\dot\alpha}_A = 
-4iu^-_i(\epsilon^{i\alpha}\bar\theta^{+\dot\alpha} + 
\theta^{+\alpha}\bar\epsilon^{i\dot\alpha}) \\ 
  \delta \theta^{i\alpha} = \epsilon^{i\alpha} & &
\delta \theta^{+\alpha, \dot\alpha} = u^+_i\epsilon^{i\alpha, 
\dot\alpha} \\ 
  & & \delta u^\pm_i = 0\ .
  \end{array}
\end{equation}
Then, given a chiral point 1 and a G-analytic point 2, we can form 
the following coordinate differences with simple transformation 
laws: 
\begin{equation}\label{chhat}
    \begin{array}{lll}
    \tilde x^{\alpha\dot\alpha}_{12} = x^{\alpha\dot\alpha}_{1_L}
- x^{\alpha\dot\alpha}_{2_A} - 4iu^-_{2i}\theta^{i\alpha}_1\; 
\bar\theta^{+\dot\alpha}_2 & \Rightarrow & \delta \tilde 
x^{\alpha\dot\alpha}_{12} = -4i\theta^{\alpha}_{12}\; 
u^-_{2i}\bar\epsilon^{i\dot\alpha} \\ 
  \theta^{\alpha}_{12} =  u^+_{2i}\theta^{i\alpha}_1 - \theta^{+\alpha}_{2} &
 \Rightarrow & \delta \theta^{\alpha}_{12} = 0\;.
  \end{array}
\end{equation}
Now, combining these two differences, one can easily construct a 
supersymmetric invariant with all the required properties of the 
propagator $\langle W(1)V^{++}(2)\rangle$ (a Lorentz scalar of 
dimension +1, chiral at point 1, G-analytic with $U(1)$ charge +2 
at point 2):
\vspace{1mm} 
\begin{center}
  \begin{picture}(0,3000)
  \drawline\gluon[\E\CENTRAL](-12000,0)[5]
  \put(\pfrontx,200){\circle*{1000}}
  \global\advance\pfrontx by -1200
  \put(\pfrontx,0)1
  \global\advance\pbackx by 500
  \put(\pbackx,0)2
  \global\advance\pbackx by 5000
  \put(\pbackx,0)
{${1\over 2g}\langle W_a(1)V_b^{++}(2)\rangle = {\delta_{ab}\over 
4i\pi^2 }(\theta_{12})^2\; \tilde x^{-2}_{12}\;.$} 
  \end{picture}
\end{center}
 \vspace{7mm}
Here $a,b$ are indices of the adjoint representation of the YM 
group. The coefficient has been fixed by finding the complex 
scalar of the SYM multiplet in $W(x,\theta)= 
i\sqrt{2}g\phi(x)+\ldots$ and in $V^{++}(x,\theta,u) = \ldots 
-i\sqrt{2}(\theta^+)^2\bar\phi(x) + 
i\sqrt{2}(\bar\theta^+)^2\phi(x)+\ldots$ and thus relating 
$\langle W(1)V^{++}(2)\rangle$ to the standard scalar propagator 
$\langle\bar\phi(1)\phi(2)\rangle = 1/4i\pi^2\; x_{12}^{-2}\;$. 

Similarly, the hypermultiplet propagator $\langle \tilde 
q^+(1)q^+(2)\rangle$ can be built out of the coordinate difference 
\footnote{This is the $N=2$ analog of the difference 
(\ref{hat4}).} 
\begin{equation}\label{hataa}
  \hat x_{12} = x_{1_A}
- x_{2_A} + {4i\over (12)}[(1^-2) \theta^+_1 \bar\theta^+_1 + 
(2^-1) \theta^+_2 \bar\theta^+_2 +  \theta^+_1 \bar\theta^+_2 + 
\theta^+_2 \bar\theta^+_1] 
\end{equation}
where $(12),(1^-2)$, etc. is a shorthand for contractions of 
harmonics, e.g., $(12)= u^{+i}_1 u^+_{2i}, \ (1^-2)= u^{-i}_1 
u^+_{2i}$. Unlike the mixed chiral-analytic one $\tilde x_{12}$ 
(\ref{chhat}), this purely G-analytic difference is invariant 
under supersymmetry, $\delta\hat x_{12} = 0$. Thus, the 
hypermultiplet propagator (a Lorentz scalar of dimension 2, 
G-analytic with $U(1)$ charges +1 at both points 1 and 2) can be 
written down as follows: 
\begin{center}
  \begin{picture}(0,3000)
  \drawline\fermion[\E\REG](-10000,0)[5000]
  \drawarrow[\E\ATTIP](\pmidx,\pmidy)
  \global\advance\pfrontx by -800
  \put(\pfrontx,0)1
  \global\advance\pbackx by 500
  \put(\pbackx,0)2
  \global\advance\pbackx by 5000
  \put(\pbackx,0)
{$\langle \tilde q_a^+(1)q_b^+(2)\rangle ={\delta_{ab}\over 
4i\pi^2} (12)\; \hat x^{-2}_{12}\;.$} 
  \end{picture}
\end{center}
 \vspace{7mm}
Once again, the coefficient has been fixed by examining the 
isodoublet scalar of the hypermultiplet, $q^+(x,\theta,u) = 
f^i(x)u^+_i + \ldots\ $.

\subsection{Building blocks}

Let us now return to the graphs in Figure 1. It is clear that the 
two topologies in Figure 1a,b can be reduced to products of 
hypermultiplet propagators and the following three-point building 
block: 

\begin{center}
  \begin{picture}(0,3000)

  \drawline\gluon[\S\CENTRAL](0,0)[4]
  \put(\gluonbackx,\gluonbacky){\circle*{1000}}
  \drawline\fermion[\W\REG](\gluonfrontx,\gluonfronty)[5000]
  \drawarrow[\E\ATTIP](\pmidx,\pmidy)
  \global\advance\pbackx by -1200
  \put(\pbackx,\pbacky) {1a}
  \drawline\fermion[\E\REG](\gluonfrontx,\gluonfronty)[5000]
  \drawarrow[\E\ATBASE](\pmidx,\pmidy)
  \global\advance\pbackx by 500
  \put(\pbackx,\pbacky) {2b}
  \global\advance\gluonbackx by 1000
  \global\advance\gluonbacky by -500
  \put(\gluonbackx,\gluonbacky) {3c}
  \global\advance\gluonfronty by 1000
  \put(\gluonfrontx,\gluonfronty) 4
  \end{picture}
  \end{center}
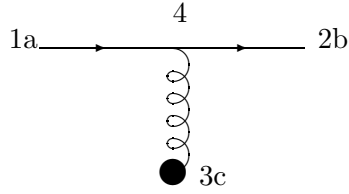
\vspace{20mm} \centerline{Figure 2} \vspace{5mm} The interaction 
point 4 is G-analytic, so one has to integrate over the G-analytic 
superspace $\int du_4 d^4x_{4_A}d^2\theta^+_4d^2\bar\theta^+_4$. 
The resulting expression is 
\begin{equation}\label{3pt}
I = {igf_{acb}\over (2\pi)^6} \int du_4 
d^4x_{4_A}d^2\theta^+_4d^2\bar\theta^+_4\ {(14)\over \hat 
x^2_{14}}\; 
 {(42)\over \hat x^2_{42}}\; {(\theta_{34})^2\over \tilde x^2_{34}}\;.
\end{equation}
It is clear that the nilpotent factor $(\theta_{34})^2$ serves as 
a Grassmann delta-function which identifies the left-handed 
G-analytic variable $\theta^{+\alpha}_4$ with the harmonic 
projection $\theta^{+\alpha}_{3/4} \equiv u^+_{4i}\theta^{\alpha 
i}_3$ of the chiral variable $\theta^{\alpha i}_3\;$. This allows 
us to immediately do the left-handed half of the Grassmann 
integral at point 4. The easiest way to do the remaining 
right-handed integration is to make use of the supersymmetry of 
the expression in (\ref{3pt}). The idea is to shift away the two 
external G-analytic variables $\theta^{+\alpha,\dot\alpha}_{1,2}$ 
by means of a finite supersymmetry transformation: 
\begin{equation}\label{epsii}
  ({\theta^{+\alpha,\dot\alpha}_{1,2}})'=\theta^{+\alpha,\dot\alpha}_{1,2}
+  u^+_{1,2i}\epsilon^{i\alpha,\dot\alpha} = 0 
\end{equation}
whose parameter is 
\begin{equation}\label{epsi}
  \epsilon^{i\alpha,\dot\alpha} =
{u^{+i}_2\over(12)}\theta^{+\alpha,\dot\alpha}_1 
 - {u^{+i}_1\over(12)}\theta^{+\alpha,\dot\alpha}_2\;.
\end{equation}
After this, the integral (\ref{3pt}) becomes 
\begin{eqnarray}
I &=& {gf_{abc}\over (2\pi)^6}\int du_4 dx_{4}d^2\bar\theta^+_4\; 
(14)(42)\;[x_{34} - 4i\theta^-_{3/4} \bar\theta^+_4]^{-2} \times 
 \\
  &&[x_{14} + 4i {(4^-1)\over (14)}\theta^+_{3/4}
\bar\theta^+_4]^{-2}  \; [x_{42} + 4i {(4^-2)\over 
(42)}\theta^+_{3/4} \bar\theta^+_4]^{-2}\nonumber 
\end{eqnarray}
where $\theta^{-\alpha}_{3/4} \equiv  u^-_{4i}\theta^{\alpha i}_3$ 
and the differences $x_{14},x_{42},x_{34}$ involve just 
$x_{{1,2,4}_A}$ and $x_{3_L}$. The next step is to perform a shift 
of the integration variable $x_4\ \rightarrow \ x_4 - 
4i\theta^-_{3/4} \bar\theta^+_4$ and to use the harmonic cyclic 
identity, e.g., $(4^-1)\theta^+_{3/4}+(14)\theta^-_{3/4} = 
\theta^+_{3/1}\;$, which leads to the following simplification of 
the integrand: 
\begin{eqnarray}
I &=& {gf_{abc}\over (2\pi)^6}\int du_4 dx_{4}d^2\bar\theta^+_4\; 
(14)(42)\; [x_{34}]^{-2} \times  \\ 
  &&[x_{14} -  {4i\over (14)}\theta^+_{3/1}
\bar\theta^+_4]^{-2}  \; [x_{42} - { 4i\over (42)}\theta^+_{3/2} 
\bar\theta^+_4]^{-2}\;. \nonumber 
\end{eqnarray}
In this form one realises that the entire dependence of the 
integrand on $\bar\theta^+_4$ can be represented as a shift of the 
external points $x_1$ and $x_2$: 
\begin{eqnarray}
I &=& {gf_{abc}\over (2\pi)^6}\int du_4d^2\bar\theta^+_4\; 
(14)(42) \times \label{expexp}\\ 
  &&\exp\left\{-{2i\over (14)}\theta^+_{3/1}\partial_1
\bar\theta^+_4 - {2i\over (24)} \theta^+_{3/2}\partial_2 
\bar\theta^+_4\right\} \; \int {d^4x_4\over x^2_{14} 
x^2_{24}x^2_{34}}\;. \nonumber 
\end{eqnarray}
Expanding the exponent in (\ref{expexp}) and doing the integral 
$\int d^2\bar\theta^+_4$ is now straightforward and the result is 
\begin{eqnarray}
 I &=&{gf_{abc}\over (2\pi)^6}\int
du_4\left[{(24)\over(14)}(\theta^+_{3/1})^2\; \square_1 + 
{(14)\over(24)}(\theta^+_{3/2})^2\; \square_2  \right. 
\label{sptm}\\ 
  &&\left. +\theta^{+}_{3/1}\theta^+_{3/2}\; 2\partial_1\cdot\partial_2
-\theta^+_{3/1}\sigma_{\mu\nu}\theta^+_{3/2}2i\partial^\mu_1\partial^\nu_2 
\right] \int {d^4x_4\over x^2_{14} x^2_{24}x^2_{34}} \nonumber 
\end{eqnarray}
where $(\sigma_{\mu\nu})_\alpha{}^\beta = {i\over 
2}(\sigma_\mu\tilde\sigma_\nu - 
\sigma_\nu\tilde\sigma_\mu)_\alpha{}^\beta$. The harmonic 
integration in the last two terms is trivial ($\int du_4\;1 =1$) 
and in the first two is done as follows, e.g., 
\begin{equation}\label{harmin}
  \int du_4\; {(24)\over(14)} =  \int du_4\;
{D^{++}_4(24^-)\over(14)} =  \int du_4\; (24^-)\delta(1,4) = 
(21^-)\;. 
\end{equation}
Here we have used the property $D^{++}_4(14)^{-1} = -\delta(1,4)$ 
of the singular harmonic distribution $1/(14)$ (see \cite{hsgr} 
for details). Finally, using the properties of the one-loop 
spacetime integral (see \cite{dav,eetal} for a discussion of such 
integrals), one easily finds 
$$
\square_1\int {d^4x_4\over x^2_{14} x^2_{24}x^2_{34}} = {4i\pi^2 
\over x^2_{12} x^2_{13}}\;,  \qquad 
\partial^{\mu}_1\partial^{\nu}_2\int {d^4x_4\over x^2_{14} x^2_{24}x^2_{34}} =
-4i\pi^2 {x^{\mu}_{13}x^{\nu}_{23}\over x^2_{12} 
x^2_{13}x^2_{23}}\;. 
$$
Putting all of this together, we obtain 
\begin{equation}
  I ={igf_{abc}\over (2\pi)^4} \left[ {\theta^+_{3/1} \theta^+_{3/2}\over
x^2_{13}x^2_{23}} - {(12)\theta^+_{3/1} \theta^-_{3/1}\over 
x^2_{12}x^2_{13}} 
 + {(12)\theta^+_{3/2} \theta^-_{3/2}\over x^2_{12}x^2_{23}} +2i
\theta^+_{3/1}\sigma_{\mu\nu} 
\theta^+_{3/2}{x^{\mu}_{13}x^{\nu}_{23}\over x^2_{12} 
x^2_{13}x^2_{23}}\right] \;. 
\end{equation}

Now, we have to recall that the above computation has been done in 
the special frame where  $\theta^+_1=\theta^+_2=0$. The way to 
obtain the result in the original frame is to perform a 
supersymmetry transformation with the parameter $\epsilon$ 
(\ref{epsi}). In the process the harmonic projections 
$\theta^+_{3/1},\theta^+_{3/2}$ give rise to the supersymmetric 
invariants $\theta_{31},\theta_{32}$ (see (\ref{chhat})) and the 
analytic-analytic difference $x_{12}$ becomes the invariant $\hat 
x_{12}$ (\ref{hataa}). Further, the harmonic projections 
$\theta^-_{3/1},\theta^-_{3/2}$ are converted into 
$\theta_{31},\theta_{32}$ (\ref{chhat}), e.g., 
$$
\theta^-_{3/1} \ \rightarrow \ {1\over (12)}\theta_{32} - 
{(21^-)\over (12)}\theta_{31} \;, 
$$
whereas the mixed chiral-analytic differences $x_{31},x_{32}$ 
become supersymmetric invariants with the help of both G-analytic 
$\theta^+_{1,2}\;$, e.g., 
\begin{equation}
  \check x_{31;2} = \tilde x_{31} + {4i\over (12)}\theta_{31}(\bar\theta^+_2
-(21^-)\bar\theta^+_1)\quad \Rightarrow \quad \delta\check 
x_{31;2} = 0 \;. 
\end{equation}
Note that despite the presence of various nilpotent terms, the 
invariants $\check x$ and $\hat x$ still satisfy the usual cyclic 
identity 
\begin{equation}\label{cycliden}
  \check x_{32;1} - \check x_{31;2} = \hat x_{12}\;.
\end{equation}

So, the building block from Figure 2 needed for the Feynman graphs 
in Figure 1a,b has the following expression: 
\begin{eqnarray}
 I &=& {gf_{abc}\over (2\pi)^4} \left\{ \theta_{31}
\theta_{32}\left[{1\over \check x^2_{31;2}\check x^2_{32;1}} - 
{1\over \hat x^2_{12}\check x^2_{31;2}}- {1\over \hat 
x^2_{12}\check x^2_{32;1}} \right] + {(21^-)(\theta_{31})^2\over 
\hat x^2_{12}\check x^2_{31;2}} + {(12^-)(\theta_{32})^2\over \hat 
x^2_{12}\check x^2_{32;1}}\right. \nonumber\\ 
  && \phantom{{igf_{abc}\over (2\pi)^4} \left\{\right.}\left.
+2i\theta_{31}\sigma_{\mu\nu} \theta_{32}{\check 
x^{\mu}_{31;2}\check x^{\nu}_{32;1}\over \hat x^2_{12} \check 
x^2_{31;2}\check x^2_{32;1}}\right\}\;. \label{bb1} 
\end{eqnarray}

The graph in Figure 1c is made out of the building block shown in 
Figure 3: 
\begin{center}
  \begin{picture}(0,3000)
\drawline\fermion[\E\REG](-5000,0)[10000] \global\advance\pfrontx 
by -1200 \put(\pfrontx,\pfronty) {1a} \global\advance\pbackx by 
500 \put(\pbackx,\pbacky) {2b} \global\advance\pmidy by -2650 
\put(\pmidx,\pmidy){\circle*{1000}} \global\advance\pmidy by -1600 
\put(\pmidx,\pmidy)3 \global\advance\pfrontx by 8300 
\drawloop\gluon[\S5](\pfrontx,\pfronty) \global\advance\loopfronty 
by 800 \put(\loopfrontx,\loopfronty) 5 
\put(\loopbackx,\loopfronty) 4 
  \end{picture}
\end{center}
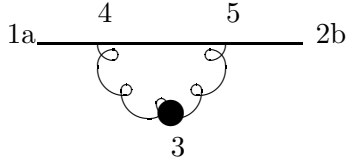
\vspace{15mm} \centerline{Figure 3} \vspace{5mm}  The computation 
is similar to that of the block in Figure 2. Firstly, one uses the 
nilpotent factors from the two SYM propagators $3\rightarrow4$ and 
$3\rightarrow5$ to do the integrals $\int d^2\theta^+_{4,5}\;$. 
Secondly, by means of a supersymmetry transformation one 
eliminates $\theta^+_{1,2}\;$. Thirdly, after shifts of the 
integration variables $x_{4,5}$ one frees all the propagators in 
the loop from any $\theta$ dependence. The result is: 
\begin{eqnarray}
 J &=& {ig^2f_{acd}f_{bcd}\over (2\pi)^{10}}\int
du_{4,5}d^4x_{4,5}d^2\bar\theta^+_4d^2\bar\theta^+_5\; 
{(14)(45)(52)\over x^2_{45}x^2_{34}x^2_{35}} \\ 
  &\times&\left[x_{14}-
{4i\over (14)}\theta^+_{3/1}\bar\theta^+_4 \right ]^{-2} 
\left[x_{25}- {4i\over (25)}\theta^+_{3/2}\bar\theta^+_5 \right 
]^{-2} \;. \nonumber 
\end{eqnarray}
This time the expansion and integration with respect to 
$\theta^+_{4,5}$ is very easy, giving rise to spacetime 
delta-functions $\delta(x_{14})$ and $\delta(x_{25})$. The 
harmonic integral is then reduced to 
\begin{equation}
  \int du_{4,5} {(45)\over (14)(52)} = - (1^-2^-)
\end{equation}
(see (\ref{harmin})). Finally, after restoring the supersymmetry 
one finds: 
\begin{equation}\label{bb2}
 J={ig^2f_{acd}f_{bcd}\over
(2\pi)^{6}}{(1^-2^-)(\theta_{31})^2(\theta_{32})^2\over \hat 
x^2_{12} \check x^2_{31;2}\check x^2_{32;1}}\;. 
\end{equation}

\subsection{Results}

At this stage what remains to do is to multiply the above building 
blocks together with the relevant hypermultiplet propagators and 
obtain the complete expressions for the five-point graphs in 
Figure 1. This involves a lot of elementary algebra, therefore we 
shall only do it in full detail in the simpler case of the 
two-loop three-point correlator 
\begin{equation}\label{3pc}
  \langle(\tilde q^+(1))^2(q^+(2))^2{1\over 4g^2}(W(3))^2\rangle\;.
\end{equation}
The corresponding graphs are shown in Figure 4: 
\begin{center}
  \begin{picture}(-1000,3000)
\put(-9000,0){\oval(10000,5500)[t]}\put(-9000,0){\oval(10000,5500)[b]} 
\put(-14000,0){\circle*{300}}\put(-4000,0){\circle*{300}} 
\put(-14800,0) 1 \put(-3500,0) 2 
\drawline\gluon[\S\CENTRAL](-9000,2750)[5] 
\put(\pmidx,\pmidy){\circle*{1000}} \global\advance\pmidx by 1200 
\put(\pmidx,\pmidy) 3 \global\advance\pbacky by -2500 
\global\Yone=\pbacky \put(\pbackx,\Yone) a 
\drawline\fermion[\E\REG](4000,0)[10000] 
\put(\pfrontx,\pfronty){\circle*{300}} 
\put(\pbackx,\pbacky){\circle*{300}} 
\put(\pmidx,\pmidy){\oval(10000,5000)[t]} \global\advance\pfrontx 
by -1100 \put(\pfrontx,\pfronty) 1 \global\advance\pbackx by 700 
\put(\pbackx,\pbacky) 2 \global\advance\pmidy by -2650 
\put(\pmidx,\pmidy){\circle*{1000}} \global\advance\pmidy by 1000 
\put(\pmidx,\Yone) b \global\advance\pmidx by -200 
\put(\pmidx,\pmidy) 3 \global\advance\pfrontx by 8300 
\drawloop\gluon[\S5](\pfrontx,\pfronty) 
\end{picture}
\end{center} 
\vspace{20mm}
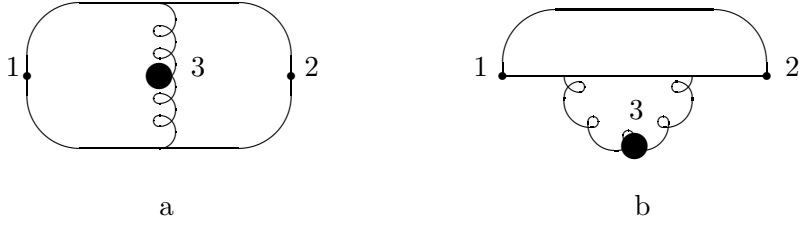
\centerline{Figure 4} 
\vspace{6mm}   
One sees that 
they are made out of the same building blocks as before. However, 
the multiplication in the case of the graph in Figure 4a is 
considerably simplified by the identification of the end-point 
$\theta$'s since the two Lorentz structures in (\ref{bb1}) (the 
scalar and the antisymmetric tensor) become orthogonal. Thus, this 
graph produces the expression (up to an overall factor) 
\begin{equation}
  {(12)(1^-2^-)(\theta_{31})^2(\theta_{32})^2\over \hat x^4_{12}
\check x^2_{31;2}\check x^2_{32;1}} \;. 
\end{equation}
Clearly, one finds exactly the same result when completing the 
building block (\ref{bb2}) from Figure 3 to the graph in Figure 4b 
by multiplying it by a hypermultiplet propagator $(12)/\hat 
x^2_{12}$. The careful computation of the group and combinatorial 
factors shows that the two contributions cancel, 
\begin{equation}\label{3pc0}
  \langle(\tilde q^+(1))^2(q^+(2))^2{1\over 4g^2}(W(3))^2\rangle = 0 \;.
\end{equation}  This confirms the
absence of quantum corrections to three-point correlators at two 
loops (see Section 2), other than possible contact terms. 
Concerning the latter, note one subtle point. The above 
multiplication of singular distributions of the type 
$1/x_{12}^2\times 1/x_{12}^2 = 1/x_{12}^4$ should be done with 
care, using a suitable regularisation scheme. The complete result 
may then contain contact terms which are lost in the formal 
manipulations presented here. 

Finally, because of the (purely algebraic) complexity of the 
calculation of the five-point correlator in Figure 1 in full 
generality, we shall do it by setting the external $\theta$'s at 
the four hypermultiplet ends to zero, $\theta^+_{1,2,3,4}=0$. The 
only surviving Grassmann variable will be the chiral one 
$\theta^{\alpha i}_5$ at the point of insertion of the SYM 
Lagrangian. In other words, we will only be interested in the 
leading component corresponding to the correlator of four 
bilinears made out of the hypermultiplet scalars with a fifth 
bilinear composed of the SYM scalars. Our aim will be to show that 
after integrating over the insertion point, one correctly 
reproduces the known result for the four-point correlator of 
\cite{eetal,eetall}. So, multiplying two building blocks of the 
type (\ref{bb1}) with distinct end points together with some 
matter propagators, adding to this the graphs made out of the 
building block (\ref{bb2}) and doing all the necessary 
permutations, one finds the following surprisingly simple result: 
\begin{eqnarray}
 &&\langle(\tilde q^+(1))^2(q^+(2))^2(\tilde
q^+(3))^2(q^+(4))^2{1\over 
4g^2}(W(5))^2\rangle_{\theta^+_{1,2,3,4}=0}  \nonumber\\ 
 &=& -g^2{f_{abc}f_{abc}\over (2\pi)^{12}}(\theta_5)^4
{x^2_{12}x^2_{34}\over x^2_{15}x^2_{25}x^2_{35}x^2_{45}} 
\left[{(12)^2(34)^2\over x^4_{12}x^4_{34}} + {(14)^2(23)^2\over 
x^4_{14}x^4_{23}} {x^2_{14}x^2_{23}\over x^2_{12}x^2_{34}}\right. 
\label{5ptinvar} \\ 
  &&+ \left. {(12)(23)(34)(41)\over
x^2_{12}x^2_{23}x^2_{34}x^2_{41}}\left({x^2_{13}x^2_{24}\over 
x^2_{12}x^2_{34}} - {x^2_{14}x^2_{23}\over x^2_{12}x^2_{34}} -1 
\right)  \right]\;. \nonumber 
\end{eqnarray}
One recognises the chiral Grassmann delta-function $(\theta_5)^4$ 
which gives the correlator the required $R$ weight of $W^2$. The 
dependence on the point $x_5$ is concentrated in a simple rational 
factor. The integration over the point of insertion $\int 
d^4\theta_5 d^4x_5$ removes $(\theta_5)^4$ and produces the 
well-known one-loop scalar box integral \cite{HV,dav} 
\begin{equation}
  \int {d^4x_5\over x^2_{15}x^2_{25}x^2_{35}x^2_{45}} =
-{i\pi^2\over x^2_{12}x^2_{34}} \Phi^{(1)}(s,t) 
\end{equation}
where 
$$
s={x^2_{14}x^2_{23}\over x^2_{12}x^2_{34}}\;,\qquad 
t={x^2_{13}x^2_{24}\over x^2_{12}x^2_{34}} 
$$
are the two conformal cross-ratios. Thus, the end result for the 
four-point correlator (or, rather, its derivative with respect to 
the coupling constant, in accordance with eq. (\ref{ourI})) is 
\begin{eqnarray}
  &&\langle(\tilde q^+(1))^2(q^+(2))^2(\tilde
q^+(3))^2(q^+(4))^2\rangle_{\theta^+_{1,2,3,4}=0} \nonumber\\ 
  &\sim& \Phi^{(1)}(s,t)
\left[{(12)^2(34)^2\over x^4_{12}x^4_{34}} + {(14)^2(23)^2\over 
x^4_{14}x^4_{23}} s + {(12)(23)(34)(41)\over 
x^2_{12}x^2_{23}x^2_{34}x^2_{41}}(t-s-1)  \right]\;. 
\label{4ptres} 
\end{eqnarray}
This is in complete agreement with the results of refs. 
\cite{eetal,eetall}.

\section{Conclusions}

The subject of this work has been the explicit construction of 
nilpotent superconformal covariants in $N=4$ SYM theory. In 
particular, we have used the superconformal Ward identities to 
find  all two- and three-point contact terms. We have also 
investigated the existence of such contact terms for more than 
three points and we have argued that such terms cannot affect the 
proof of the non-renormalisation of two- and three-point function 
given in \cite {ehw}. We have also argued, subject to some 
assumptions, that all contact terms  arise from the addition of a 
finite local counterterm to the effective action, namely the  
superconformal action. 
\par
The explicit contact terms which were found in reference 
\cite{skenderis} followed from non-contact contributions as a 
result of the Ward identities. Such contact terms are 
automatically encoded in the approach advocated in the works by 
the authors of this paper. Neither in our attempts at constructing them 
nor in our application of that formula to an explicit two-loop 
calculation have we found evidence for the existence of contact 
terms of the more malignant type that would invalidate the 
non-renormalisation theorem for two and three point functions 
given  in \cite {ehw}. 

We have also carried out a two-loop calculation in $N=2$ harmonic
superspace and as a result  have been able to
prove explicitly the existence of a five-point
nilpotent $N=2$ superconformal covariant which in turn strongly
suggests the 
existence of a corresponding
$N=4$ covariant. 
If the multiplication of the building blocks
in this calculation is done in
full detail (i.e., without setting $\theta^+_{1,2,3,4}=0$) one
arrives at an explicit expression for this five-point superconformal invariant which is of the
type discussed in \cite{ehw}. As first suggested in
\cite{ken1, ken2} such new invariants must exist if the $N=4$
harmonic superspace is to be consistent with the known facts about
the Green's functions in $N=4$ Yang-Mills theory.  These
new invariants do
however,  have  the other general properties postulated in
\cite{hw1,ehw}. These properties are 
harmonic analyticity $D^{++}G =0$, which is
evident from (\ref{5ptinvar}), and superconformal invariance
(which follows from the $N=4$ SYM context of the calculation).

As a byproduct of this investigation, we have achieved
a further significant simplification in the calculation of
the two-loop four-point correlators of gauge invariant
operators first considered in \cite{eetal,grps}.
In the direct calculation of the four-point
correlator carried out in ref. \cite{eetal} the fact that all the
three harmonic structures in (\ref{4ptres}) have the same
non-trivial dependence on the conformal cross ratios was not
obvious at all. An indirect argument based on conformal
supersymmetry and harmonic analyticity allowed us to establish
this relationship in \cite{eetall}. The present calculation
reproduces it directly, due to the remarkably simple structure of
the five-point correlator. Moreover, in the variant of this
calculation presented here the appearance of any two-loop integrals
was completely avoided. This was
made possible by the fact
that for the two basic building blocks
of our two-loop diagrams, depicted in figs. 2 and 3, the
integration over the internal point can be performed
trivially using supersymmetry. This fact is
of independent interest, and may possibly lead to simplifications
in other contexts.

\vspace{20pt} {\bf Acknowledgements:} We are indebted to M. 
Bianchi, F. Delduc, E. D'Hoker, E. Ivanov, R. Stora and S. Theisen 
for many useful discussions. This work was supported in part by 
the British-French scientific programme Alliance (project 98074) 
and by the EU network on Integrability, non-perturbative effects, 
and symmetry in quantum field theory (FMRX-CT96-0012).

\end{document}